1# From Fermi Arcs to the Nodal Metal: Scaling of the Pseudogap with Doping and Temperature

A. Kanigel[1,2], M. R. Norman[2], M. Randeria[3], U. Chatterjee[1,2], S. Suoma[1], A. Kaminski[4], H.M. Fretwell[4], S. Rosenkranz[2], M. Shi[1], T. Sato[5], T. Takahashi[5], Z. Z. Li[6], H. Raffy[6], K. Kadowaki[7], D. Hinks[2], L. Ozyuzer[2], and J.C. Campuzano[1,2]

[1]Department of Physics, University of Illinois at Chicago, Chicago IL 60607, USA,

[2]Materials Science Division, Argonne National Laboratory, Argonne IL 60439, USA,

[3]Department of Physics, Ohio State University, Columbus, OH 43210, USA,

[4]Ames Laboratory and Department of Physics and Astronomy, Iowa State University, Ames, IA 50011, USA,

[5]Department of Physics, Tohoku University, 980-8578 Sendai, Japan,

[6]Laboratorie de Physique des Solides, Universite Paris-Sud, 91405 Orsay Cedex, France,

[7]Institute of Materials Science, University of Tsukuba, Ibaraki 305-3573, Japan
**The pseudogap phase in the cuprates is a most unusual state of matter[1-4]: it is a metal, but its Fermi surface is broken up into disconnected segments known as Fermi arcs[5]. Using angle resolved photoemission spectroscopy, we show that the anisotropy of the pseudogap in momentum space and the resulting arcs depend only on the ratio $T/T^*(x)$, where $T^*(x)$ is the temperature below which the pseudogap first develops at a given hole doping $x$. In particular, the arcs collapse linearly with $T/T^*$ and extrapolate to zero extent as $T \to 0$. This suggests that the $T = 0$ pseudogap state is a nodal liquid, a strange metallic state whose gapless excitations are located only at points in momentum space, just as in a d-wave superconductor.**

In Figs. 1a and 1b we present data for a slightly underdoped sample of $Bi_2Sr_2CaCu_2O_{8+\delta}$ (Bi2212) with a $T_c$ = 90 K, for (a) the superconducting state at 40K, and (b) the pseudogap phase at 140K. Shown are energy distribution curves (EDCs) at the Fermi momentum $k_F$, which have been symmetrized[5] to remove the effects of the Fermi function on the spectra. $k_F$ is determined by the minimum separation between the peaks in the symmetrized spectra along each momentum cut. Fifteen momentum cuts were measured, shown in Fig. 1d. Details of the symmetrization procedure are explained in the Methods section. The difference between the spectra in the two states is apparent: sharp spectral peaks are present in the superconducting state indicating long-lived excitations, and the superconducting gap vanishes only at points in the Brillouin zone, known as nodes. On the other hand, the spectra in the pseudogap phase are much broader, indicating short-lived excitations. Although a pseudogap is seen in cuts 1 through 7, a substantial part of the Fermi surface, cuts 8 through 15, show spectra peaked at the Fermi energy, indicating a Fermi arc of gapless excitations.

The gap size can be estimated as half the peak to peak separation in energy. A more quantitative estimate is obtained by using a simple phenomenological function to describe the spectral lineshapes[6] (described in the Methods section), shown as black lines in Fig. 1**a** and 1**b**. In Fig. 1**c** we show the gap size from these fits as a function of the Fermi surface angle (defined in 1**d**). Note the abrupt opening of the pseudogap (red points) at the end of the arc compared to the d-wave shape of the superconducting gap (black points). In Fig. 1**e** we show the symmetrized EDCs at $k_F$ for a very underdoped sample with $T_c$ = 25K, measured at 55 K (Ca-rich Bi2212, $Bi_{2.1}Sr_{1.4}Ca_{1.5}Cu_2O_{8+\delta}$, see Ref. 7). For this doping, there is a very short Fermi arc, much shorter than the one seen in Figs. 1**b**, 1**c**, for the less underdoped $T_c$ = 90 K sample.

In Figs. 2**a** and 2**b** we show two spectral intensity maps, taken at 110K and 200K, at the Fermi energy for a $T_c$ = 70 K sample. The solid line is from a tight binding fit to the normal state Fermi surface[8], and the red points represent the experimental Fermi momenta. One can clearly see the gapless Fermi arc (high intensity region), which is longer at 200K than at 110K. Fig. 2**c** shows the angular anisotropy of the gap for these two temperatures from the fits (as in Fig. 1**c**). Indeed, at 200K, the onset of the gap at the end of the Fermi arc is steeper than at 110K, and the arc is longer. Note that the gap size remains roughly constant in the straight section of the Fermi surface near the anti-node. In this region, the Fermi surface is essentially parallel to the Brillouin zone axis (Fig. 1**d**).

We now discuss our most important finding. As shown in Figs. 1 and 2, the anisotropy of the pseudogap around the Fermi surface is temperature and doping dependent. Despite this, we find the rather remarkable result that the momentum dependence of the gaps from samples with different temperatures and different doping values can be scaled by defining a reduced temperature $t = T/T^*(x)$ and by normalizing the gap by its value at the anti-node. To demonstrate this scaling, we show six data sets in Fig. 3 with different temperatures and doping, but which are divided into two groups, one with $t = 0.9$ and the other with $t = 0.45$. For comparison, we show the angular anisotropy of the d-wave superconducting gap (dashed line). It is well known[9] that the magnitude of the pseudogap at the anti-node tracks $T^*$ as a function of $x$. The surprising result we find here is that the entire momentum and temperature dependence of the normalized pseudogap $\Delta(\phi)/\Delta(0)$ only depends on $T/T^*(x)$, while the $T_c$ of the sample does not play a role. We note that scaling with $T^*$ has been observed for susceptibility and transport data[10,11,12].

However, the gap size alone does not provide a full description of the low energy excitations in the pseudogap state, for which we also need to consider the temperature dependence of the intensities. In the inset to Fig. 2**c** we show symmetrized EDCs for a $T_c$ = 89 K film at the anti-node. As the temperature increases, the gap does not change, but rather there is a filling-in of low energy spectral weight[3,5,6], with the pseudogap effectively disappearing for $T>T^*$. We quantify the loss of spectral weight in the pseudogap region as $L(\phi) = [1 - I(0,\phi)/I(\Delta,\phi)]$, where $I(0,\phi)$ is the symmetrized intensity at the Fermi energy and at an angle $\phi$ (defined in Fig. 1**d**), and $I(\Delta,\phi)$ is that at the gap energy at the same angle. In Fig. 4**a** we plot $L(\phi)$ for all values of doping and temperature, as a function of the Fermi surface angle $\phi$. Points corresponding to $L(\phi) = 0$ have no loss of low energy spectral weight, and hence correspond to the Fermi arc. $L(\phi)$ as a function of $t$ exhibits the general trend of the pseudogap (though we note that the temperature dependence of the filling of the gap does depend on $T_c$ [6]). It also permits us to precisely determine the lengths of the Fermi arcs. We now discuss the scaling of the Fermi arcs.

We find from Fig. 4 that the length of the Fermi arc increases with temperature up to the straight sections of the Fermi surface shown in Fig. 1**d**. As clearly seen from Fig. 3, the Fermi surface in the straight section is gapped for all $T<T^*(x)$. As a consequence, as $t$ approaches 1, the variation of the gap becomes more abrupt, i.e. the transition from the gapless Fermi arc to the pseudogap takes place over a shorter range of momentum. In the straight section of the Fermi surface, the gap does not close, but instead fills in as $T$ approaches $T^{*}$[6]. Since the gap is relatively momentum independent in the straight section, the arc appears to suddenly expand to the full Fermi surface as $T$ passes through $T^*$.

Our main conclusion is shown in Fig. 4**b**, where we find that the length of the Fermi arc is a linear function of $t$ (in the range $0 < t = T/T^*(x) < 1$), and the arc extrapolates to a point node as $t \to 0$. Above $T^*$, the arc covers the full length of the Fermi surface. As the temperature is reduced below $T^*$, there is an abrupt opening of the gap over the straight part of the Fermi surface, as discussed above. As $t$ is lowered further, the length of the arc decreases *linearly* all the way from $t = 1.0$ to $t = 0.1$. The limit of $t = 0$ is presently inaccessible, since the Bi2212 samples available are all superconducting at sufficiently low temperatures. However, a simple extrapolation of the linear $t$ dependence of the arc length implies that the arcs would shrink to point nodes in a zero temperature pseudogap state (we emphasize that all the data presented in Fig. 4**b** are in the pseudogap state above $T_c$). We note in passing that the crossing of the arc with the $(0,0)$-$(\pi,\pi)$ line does not move to $(\pi/2,\pi/2)$ on decreasing $t$, so the Fermi surface is not shrinking but instead the arcs are gapped out.

Our results indicate that the low temperature limit of the momentum dependence of the pseudogap is identical to that of a d-wave superconductor. This agrees with many aspects of the "nodal liquid" model suggested by Balents, Fisher and Nayak[13], in which quantum disorder in a d-wave superconductor leads to a state with no long-range phase coherence, but with the same low energy excitation spectrum. The similarity of the zero-temperature electronic structure of the superconducting and pseudogap states is also in accord with thermal conductivity experiments carried out in very underdoped $YBa_2Cu_3O_{6+\delta}$ samples at milli-Kelvin temperatures[14].

On the other hand, the limit of $t \rightarrow 1$ presents a somewhat different picture, one with a gap located only in the straight parts of the underlying Fermi surface. This type of momentum dependence is suggestive of an instability associated with a finite $q$-vector that spans the distance between the straight segments of the Fermi surface[15]. However, several experimental observations argue against an ordering scenario: 1) If charge ordering is present, one would expect to find "shadow" bands associated with such ordering, that is, images of the arcs displaced by the charge ordering wavevector $q$. We have extensively looked for these and have not found them. 2) If an ordering vector $q$ is present, the energy gap would be tied to the new zone boundary set by $q$. We have found no evidence for such a boundary. In particular, all energy gaps we observe are *always* tied to the Fermi surface and to the chemical potential (which we have verified by dividing EDCs by the Fermi function), which is only consistent with a $q = 0$ instability. 3) If the vector $q$ spans the straight sections of the Fermi surface, it will not span the curved sections. But we find the curved sections to also be gapped at low $t$. That is, the lengths of the Fermi arcs would not be temperature dependent, as we find here.

We note that the more interesting $t \rightarrow 0$ limit does not give any evidence for qualitatively different gaps on the straight and curved parts of the Fermi surface. The green data points in Fig. 4**a** show a smooth d-wave-like angular dependence, except for the very small Fermi arc near the node. Taken together, all the data suggest that the pseudogap in the $t \rightarrow 0$ limit will simply look like a d-wave gap, albeit with a lack of long range superconducting order.

In conclusion, it is widely believed that understanding the pseudogap phase, which lies between the Mott insulating and the superconducting phases in the cuprates, is the key to understanding high temperature superconductivity in the cuprates. Our results on the scaling of the pseudogap with $T/T^*(x)$ give completely new insight on how this mysterious state of matter evolves with temperature and doping and present a challenge to existing theories. This scaling suggests that the ground state of the pseudogap is the nodal metal.

**Methods**

**Samples and measurements**: We measured six underdoped $Bi_2Sr_2CaCu_2O_{8+\delta}$ samples with different doping levels ($T_c$ between 70 and 90K). Two of them are single crystals grown by the floating zone method, the other four are thin films grown by RF sputtering. We also measured a highly underdoped single crystal sample[7] ($Bi_{2.1}Sr_{1.4}Ca_{1.5}Cu_2O_{8+\delta}$) with a $T_c$ of 25K. All samples were mounted with the $(0,0) \rightarrow (\pi,0)$ (i.e. the bond) direction parallel to the photon polarization, and cleaved in situ at pressures less than $2 \times 10^{-11}$ Torr. Measurements were carried out at the Synchrotron Radiation Center in Madison, Wisconsin on the U1 and PGM undulator beamlines, using Scienta electron analyzers models R4000, SES2002, SES200, or SES50, with energy resolution ranging from 16 meV to 30 meV, and a momentum resolution of 0.01-0.02 Å$^{-1}$ for a photon energy of 22 eV. The samples were in contact with gold during the experiment for the purpose of accurately referencing the chemical potential. We measured cuts in momentum space perpendicular to Γ-M, in the Y quadrant of the Brillouin zone so as to minimize complications due to the superlattice[16]. The cuts cover the region from the node to the anti-node of the d-wave superconducting gap, as shown in Fig. 1**d**. We have checked that our results are independent of the incident photon energy.

**Data analysis**: The first step is to accurately identify the Fermi momentum $k_F$ in the presence of a pseudogap. We do this by finding the *k*-point with the minimum gap along a given cut in the symmetrized data, $I(\omega) = I_{ARPES}(\omega) + I_{ARPES}(-\omega)$, where $I_{ARPES}$ is the measured angle resolved photoemission intensity. (We note that one cannot determine $k_F$ by looking at the peak of the momentum distribution curves in the presence of an energy gap[17].) The symmetrized intensity at $k_F$ effectively divides out the Fermi function[5], and thus provides a clear view of the gap. Although this procedure assumes that the unoccupied part of the excitation spectrum is the same as that of the occupied part over the small energy range of the gap, we have obtained similar results by simply dividing the data by the resolution broadened Fermi function. The gap may now be obtained from one half the peak-to-peak distance. For a more precise determination of the gap, we fit the spectra using a simple phenomenological self-energy[6]: $\Sigma(\mathbf{k},\omega) = -i\Gamma_1 + \frac{\Delta^2}{\omega + i\Gamma_0}$, where $\Gamma_1$ is the single-particle scattering rate, $\Delta$ the energy gap, and $\Gamma_0$ an inelastic broadening parameter that describes the filling in of the subgap spectral weight ($\Gamma_0$ is zero in the superconducting state). Note that the spectral intensity is proportional to the imaginary part of $G$, where $G^{-1} = \omega - \Sigma$ at $k_F$. This phenomenological expression gives an accurate representation of the spectra for all values of $k_F$, temperature, and doping (as can be seen in Fig. 1).

***T\* determination***: Several different methods were used to determine the pseudogap temperature $T^*$. 1) For three of the samples with lower $T^*$, we measured $T^*$ directly from ARPES by observing the appearance of gapless excitations. 2) For the ones with higher $T^*$, the pseudogap temperature was determined by the established relation[9] between the superconducting gap size and $T^*$ in Bi2212. 3) An independent method for determining $T^*$ is to use the inelastic broadening parameter $\Gamma_0(T)$ of the phenomenological self-energy[6]. It is known[6] that $\Gamma_0(T)$ increases linearly with $T$, and $T^*$ is determined by $\Gamma_0(T^*) = \Delta(\phi = 0)$. 4) Yet another independent way to obtain $T^*$ is from the $T$-dependence of $L(\phi) = \left[1 - I(0,\phi)/I(\Delta_\phi,\phi)\right]$ at the antinode $\phi = 0$. We find that $L(\phi = 0)$ is a linearly decreasing function of $T$, and extrapolates to 0 at $T = T^*$. All of the above methods give consistent values of $T^*$ (see online supporting material for further details).

---

1. Ding, H. *et al.* Spectroscopic evidence for a pseudogap in the normal state of underdoped high-$T_c$ superconductors. *Nature* **382,** 51–54 (1996).

2. Loeser, A. G. *et al.* Excitation gap in the normal state of underdoped $Bi_2Sr_2CaCu_2O_{8+\delta}$. *Science* **273,** 325–329 (1996).

3. Timusk, T. and Statt, B. The pseudogap in high-temperature superconductors: an experimental survey. *Rep. Prog. Phys.* **62,** 61-122 (1999).

4. Randeria, M. Precursor Pairing Correlations and Pseudogaps, in Proceedings of the International School of Physics "Enrico Fermi" on Conventional and High Temperature Superconductors, ed. G. Iadonisi, J. R. Schrieffer, and M. L. Chiafalo, (IOS Press, Amsterdam, 1998), p. 53 - 75; cond-mat/9710223.

5. Norman, M.R. *et al.* Destruction of the Fermi surface in underdoped high-$T_c$ superconductors. *Nature* **392**, 157-160 (1998).

6. Norman, M.R., Randeria, M., Ding, H. and Campuzano, J.C. Phenomenology of the low-energy spectral function in high-$T_c$ superconductors. *Phys. Rev. B* **57**, R11093-R11096 (1998).

7. Ozyuzer, L., *et al.* Probing the Phase Diagram of $Bi_2Sr_2CaCu_2O_{8+\delta}$ With Tunneling Spectroscopy. *IEEE Trans. Appl. Supercond.* **13**, 893-896 (2003).

8. Norman, M.R., Randeria, M., Ding, H. and Campuzano, J.C. Phenomenological models for the gap anisotropy of $Bi_2Sr_2CaCu_2O_8$ as measured by angle-resolved photoemission spectroscopy. *Phys. Rev. B* **52**, 615–622 (1995).

9. Campuzano, J.C. *et al.* Electronic spectra and their relation to the ($\pi,\pi$) collective mode in high-$T_c$ superconductors. *Phys. Rev. Lett.* **83**, 3709-3712 (1999).

10. Nakano, T. *et al.* Magnetic properties and electronic conduction of superconducting $La_{2-x}Sr_xCuO_4$. *Phys. Rev. B* **49**, 16000-16008 (1994).

11. Wuyts, B. *et al*. Resistivity and Hall effect of metallic oxygen-deficient $YBa_2Cu_3O_x$ films in the normal state. *Phys. Rev. B,* **53** 9418-9432 (1996).

12. Konstantinovic, Z., Li, Z.Z., and Raffy, H. Normal State transport properties of single and double layered $Bi_2Sr_2Ca_{n-1}Cu_nO_y$ thin films and the pseudogap effect. *Physica C* **341-348**, 859-862 (2000).

13. Balents, L., Fisher, M.P.A. and Nayak, C. Nodal liquid theory of the pseudo-gap phase of high-$T_c$ superconductors. *Intl. J. Mod. Phys. B* **12**, 1033-1068 (1998).

14. Sutherland, M. *et al.* Delocalized fermions in underdoped cuprate superconductors. *Phys. Rev. Lett*. **94**, 147004 (2005).

15. Shen, K.M. *et al.* Nodal quasiparticles and antinodal charge ordering in $Ca_{2-x}Na_xCuO_2Cl_2$. *Science* **307**, 901-904 (2005).

16. Ding, H. *et al.* Electronic excitations in $Bi_2Sr_2CaCu_2O_{8+\delta}$: Fermi surface, dispersion, and absence of bilayer splitting. *Phys. Rev. Lett.* **76**, 1533-1536 (1996).

17. Norman, M.R., Eschrig, M., Kaminski, A. and Campuzano, J.C. Momentum distribution curves in the superconducting state. *Phys. Rev. B* **64**, 184508 (2001).

18. Kaminski, A, *et al.* Identifying the background signal in angle-resolved photoemission spectra of high-temperature cuprate superconductors. *Phys. Rev. B* **69**, 212509 (2004).

**Supplementary Information** accompanies the paper on **www.nature.com/nphys**.

This work was supported by NSF DMR-0305253, the US DOE, Office of Science, under Contracts No. W-31-109-ENG-38 (ANL) and W-7405-Eng-82 (Ames), and the MEXT of Japan. The Synchrotron Radiation Center is supported by NSF DMR-0084402.

Correspondence and requests for materials should be addressed to J.C.C. (e-mail: jcc@uic.edu).

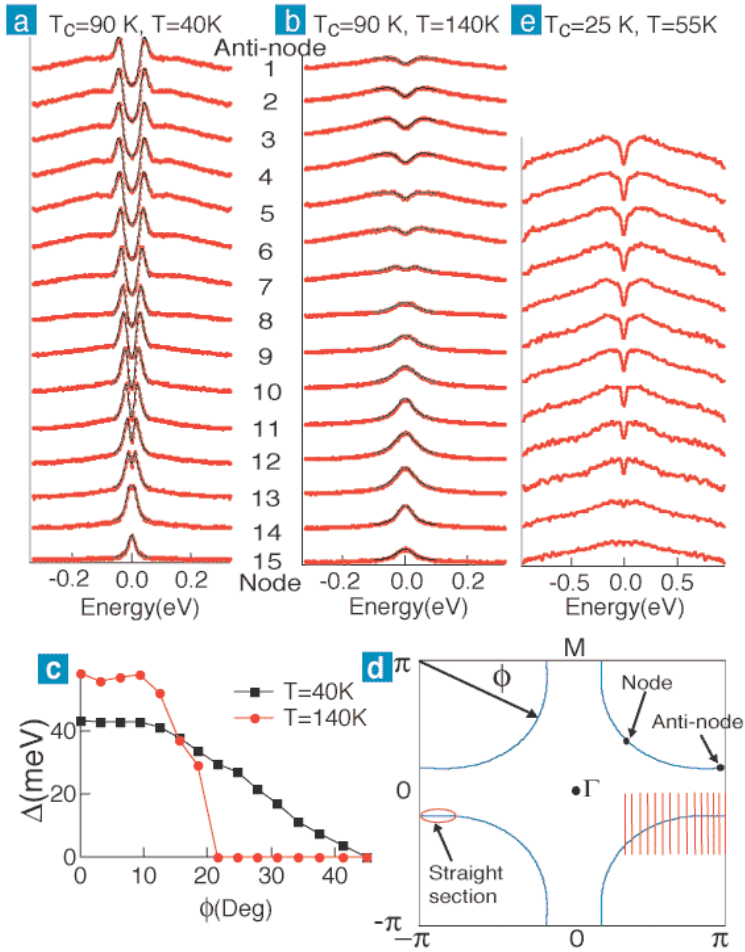

**Figure 1**: Symmetrized EDCs for underdoped samples along the Fermi surface. **a** $T_C$ = 90 K sample in the superconducting state at $T$ = 40 K, and **b** the same sample in the pseudogap phase at $T$ = 140 K. The bottom EDC is at the node, while the top is at the anti-node, as defined in **d**. **c** Variation of the gap around the Fermi surface extracted from **a** and **b**. **d** Location of the momentum cuts (red lines), Fermi surface (blue curves), and special points (node and anti-node) in the zone. **e** Symmetrized EDCs for a very underdoped, $T_c$ = 25 K, sample (corresponding to $k_F$ points 4 through 15), measured at 55 K in the pseudogap state. For this sample, the spectral weight is much reduced relative to higher doping values. We therefore removed the extrinsic background[18].

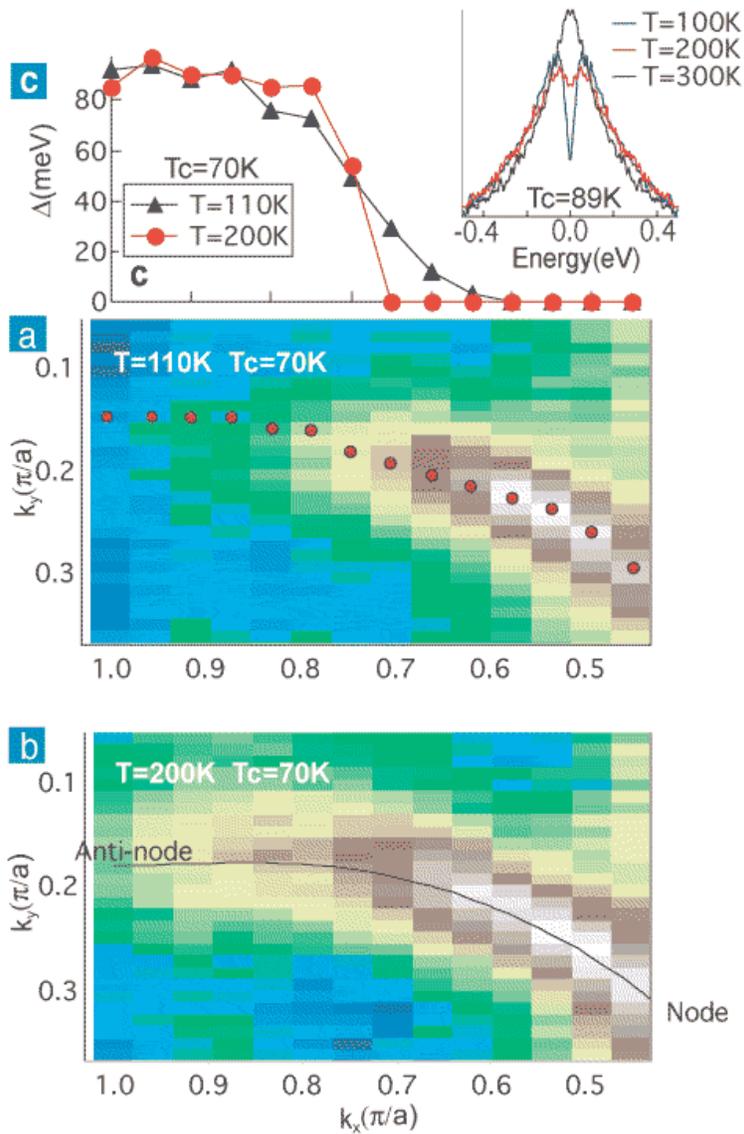

**Figure 2**: Intensity maps at the Fermi energy for an underdoped $T_c$ = 70K sample. **a** at 110K and **b** at 200K (red points are measured $k_F$ values). **c** Angular anisotropy of the gap along the Fermi surface from the data of **a** and **b**. The inset to **c** shows the temperature variation of the symmetrized EDCs at the anti-node for a $T_c$ = 89K sample, with the EDC at 300K in the gapless normal state above $T^*$.

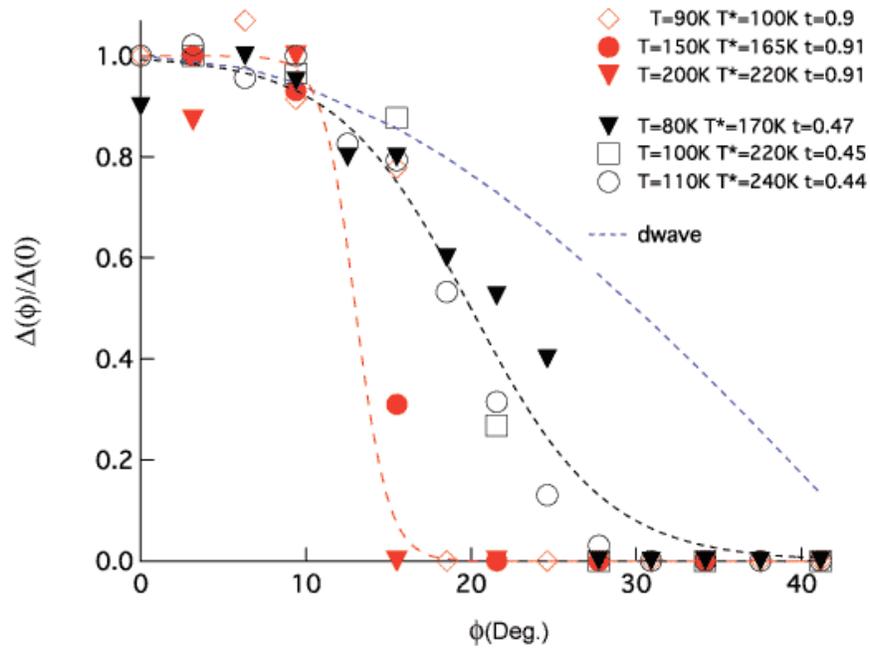

**Figure 3**: Gap size (normalized to its value at the anti-node) as a function of Fermi surface angle (defined in Fig. 1**d**) for two particular reduced temperatures: $t = T / T^* = 0.45$ (black) and 0.9 (red). $\Delta$ is set to zero if the spectrum is gapless. The data are from several samples with different doping levels, as characterized by $T^*$. The dashed lines are guides to the eye.

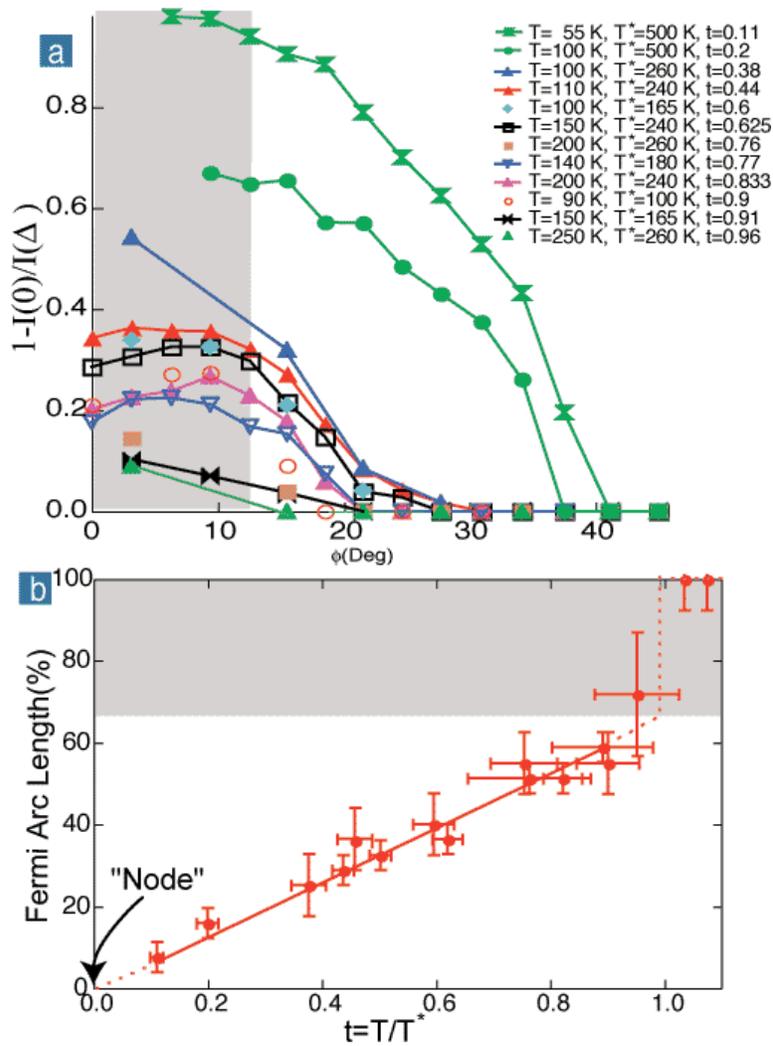

**Figure 4**: Loss of low energy spectral weight $L(\phi)$ and scaling of arc lengths with $T/T^*$. **a** $L(\phi) = [1 - I(0,\phi)/I(\Delta,\phi)]$, where $I(\Delta,\phi)$ is the symmetrized intensity at the gap energy at the Fermi point labelled by $\phi$ and $I(0,\phi)$ is that at the Fermi energy at the same point. Different symbols correspond to various $t$. The grey area represents the straight section of the Fermi surface. **b** Variation of the arc length with respect to the reduced temperature $t = T/T^*$. On the y axis, 0% is the node and 100% the anti-node (see 1d). The variation of the arc length below $T^*$ is linear in $t$. Above $T^*$, the gap in the straight section disappears, and the full Fermi surface is recovered.